\documentstyle[12pt]{article}
\newcommand{\Frac}[2]{\frac{\displaystyle #1}{\displaystyle #2}}
\newcommand{\kpiggp}{$K^+ \rightarrow \pi^+ \gamma \gamma \;  $ }
\newcommand{\kpiggpe}{K^+ \rightarrow \pi^+ \gamma \gamma \;   }
\newcommand{\kpiggn}{$K_L \rightarrow \pi^{\circ} \gamma \gamma \; $ }

\begin{document}
\pagestyle{empty}
\begin{titlepage}
\begin{flushright}
INFNNA-IV-96/12\\
DSFNA-IV-96/12\\
\today
\end{flushright}
\begin{center}
\vspace*{3cm} 
{\LARGE \bf $^*$ Unitarity and vector meson contributions  \\ 
\vspace*{0.5cm} to $K^+ \rightarrow \pi^+ \gamma \gamma$ \\ }
\vspace*{2.5cm}
{\large \bf G. D'Ambrosio} $ \; \; $ and $ \; \; $
{\large \bf  J. Portol\'es }
\vspace*{0.5cm} \\
Istituto Nazionale di Fisica Nucleare, Sezione di Napoli \\
Dipartamento di Scienze Fisiche, Universit\`a di Napoli \\
I-80125 Napoli, Italy \\ 
\vspace*{2.5cm} 

\begin{abstract}
We compute the one-loop unitarity corrections ${\cal O}(p^6)$ from 
$K^+ \rightarrow \pi^+ \pi^+ \pi^-$ to \kpiggp and we find that they
are relevant, increasing the leading order prediction for the width
in a $30-40 \%$. The contributions of local  ${\cal O}(p^6)$
amplitudes, generated
by vector meson exchange, are discussed in several models and we conclude 
that the vector resonance contribution should be negligible  compared 
to the unitarity corrections.
\end{abstract}
\end{center}
\vfill
\noindent * Work supported in part
by HCM, EEC--Contract No. CHRX--CT920026 (EURODA$\Phi$NE) 
\end{titlepage}
\newpage
\pagestyle{plain}
\pagenumbering{arabic}

\section{Introduction}
\hspace*{0.5cm}
The phenomenology of radiative non--leptonic kaon decays provides crucial
tests for the ability of Chiral Perturbation Theory ($\chi PT$) 
\cite{WG79} to explain weak low--energy processes and interesting 
possibilities to study CP violation in these channels. $\chi PT$
is a natural framework that embodies together an effective theory 
(satisfying
the  basic chiral symmetry of QCD) and a perturbative Feynman--Dyson 
expansion. Its success in the study of radiative non--leptonic kaon 
decays has been remarkable (see \cite{rev,DE95} and references therein).
Of course there are still open problems, but upcoming
experiments should improve our phenomenological knowledge of these decays,
in particular on $K \rightarrow \pi \gamma \gamma$, $K
\rightarrow \pi \ell^+ \ell^-$, $K_L \rightarrow \ell^+ \ell^-$ or  $K_L
\rightarrow \gamma \ell^+ \ell^-$.
\par 
\kpiggn is a very interesting channel by itself as a $\chi PT$ test and  
also in order to establish the
relative role of the CP conserving contribution to  $K_L
\rightarrow \pi^\circ e^+ e^-$ versus the CP violating contributions
\cite{rev,DE95,DH87,EPR88,FR,SE88,EP90}. \kpiggn 
has no ${\cal O}(p^2)$ tree level contribution since the external 
particles are
neutral. For this same reason there are no ${\cal O}(p^4)$ counterterms 
and the
chiral meson  loops are finite at this order \cite{DE87}. 
Consequently $\chi PT$ 
gives an unambiguous prediction. Furthermore only an helicity suppressed
amplitude for $K_L\rightarrow \pi^0 e^+ e^-$ is generated, which is
small compared to the CP violating ones. At higher orders in $\chi PT$ 
a new invariant amplitude, generating 
an unsuppressed helicity  amplitude to $K_L\rightarrow \pi^0 e^+ e^-$, 
appears. 
As we will see in detail later there is a specific kinematical region
in the diphoton invariant mass spectra of the $K \rightarrow \pi \gamma 
\gamma$
processes where  only this last amplitude contributes. This is the region 
where
both photons are nearly collinear, i.e. 
at small diphoton invariant mass.
The experimental results showed \cite{NA92} that while the ${\cal O}(p^4)$
predicted spectrum looked in  good agreement with the experiment,
finding no contribution at small
diphoton invariant mass, the predicted branching
ratio was  underestimated by a factor slightly bigger than two, two sigmas
away from the experimental result.
\par
This fact
prompted to several authors to consider higher chiral order corrections:
i) Vector Meson Exchange contributions to the local ${\cal
O}(p^6)$ amplitude \cite{SE88,EP90}, 
ii) Unitarity corrections from $K_L \rightarrow
\pi^{\circ} \pi^+ \pi^-$ \cite{CD93,CE93} and iii) 
Complete unitarization of the $\pi \pi$
intermediate states through a Khuri-Treiman treatment and inclusion of the
experimental $\gamma \gamma \rightarrow \pi^{\circ} \pi^{\circ}$ amplitude
\cite{KH94}. It has to be emphasized that  these higher order corrections 
do not
represent a complete chiral  order contribution. 
The size of the first correction is controversial \cite{SE88,EP90},
however a large contribution by itself is excluded by the experimental
spectrum.
The second contribution shows that  unitarity corrections of 
$K_L \rightarrow \pi^{\circ} \pi^+ \pi^-$ to \kpiggn increase $20-30\%$ 
the leading order amplitude and then have to 
be taken into account \cite{CD93,CE93}. 
When both contributions are added it was  possible to fit 
both the width and the spectrum  \cite{CE93}. 
The third correction increases the branching
ratio by an additional $10-20 \%$.
\par
The charged channel \kpiggp is experimentally less known (only
an upper limit on the branching ratio exists \cite{PDG94}),
but measurements with good precision are foreseen in the near future
\cite{DE95,FRA95}. The leading one--loop
${\cal O}(p^4)$ result has been computed \cite{EPR88} and 
depends upon unknown weak local amplitudes which however could give
important informations on vector meson weak interactions.  
Indeed while at ${\cal O}(p^4)$ it is well known that the finite part of the 
counterterms in the strong $\chi PT$ lagrangian is saturated by the
spectrum of lightest resonances (vector, axial--vector, scalar and 
pseudoscalar) \cite{EG89}, the situation in the ${\cal O}(p^4)$ weak 
lagrangian is less clear due to our ignorance on the weak couplings of vector
mesons, however there are predictive models to test \cite{EK93}.
\par
We have evaluated the $K^+ \rightarrow \pi^+ \pi^+ \pi^-$  unitarity
contribution to \kpiggp with the same procedure used to compute the $K_L
\rightarrow \pi^{\circ} \pi^+ \pi^-$ unitarity corrections to \kpiggn
\cite{CD93}. The  dispersive contribution is  unambiguously computed up to a
polynomial piece which is  reabsorbed (together with the  loop divergences) 
in the counterterms. Moreover 
various  vector dominance exchange models are studied to saturate the bulk
of the counterterms as in \kpiggn. Then we can give a definite prediction for
the kinematical region of nearly collinear photons where the unknown 
contribution of the ${\cal O}(p^4)$ local amplitude is negligible.
The experimental study of \kpiggp in the DA$\Phi$NE  $\Phi$--factory 
\cite{DE95}  and other experiments \cite{FRA95} is going to improve the
phenomenology of $K \rightarrow \pi \gamma \gamma$ and will allow us to 
analyze both channels in a  correlated way.

\section{$K \rightarrow \pi \gamma \gamma$ amplitudes}
\hspace*{0.5cm}
The general amplitude for $K \rightarrow \pi \gamma \gamma$ is given 
by 
\begin{equation}
M\, ( \, K (p) \rightarrow \pi (p_3) \, \gamma (q_1,\epsilon_1) \, 
\gamma (q_2,\epsilon_2) \, ) \; = \; {\epsilon_1}_{\mu} {\epsilon_2}_{\nu}
\, M^{\mu \nu} (p,q_1,q_2)
\label{eq:mee}
\end{equation}
where $\epsilon_1$,$\epsilon_2$ are the photon polarizations, and 
$M^{\mu \nu}$ has four invariant amplitudes
\begin{eqnarray}
M^{\mu \nu}  & = & \Frac{A(z,y)}{m_K^2} \, ( q_2^{\mu} \,  q_1^{\nu} \, - \, 
q_1 \cdot q_2 g^{\mu \nu} ) \;
 + \; \Frac{2\, B(z,y)}{m_K^4} \, ( -p \cdot q_1 \, p \cdot q_2 \, 
g^{\mu \nu}
\, - \, q_1 \cdot q_2 \, p^{\mu} p^{\nu}  \, \nonumber \\
& & \qquad \qquad \qquad \qquad \qquad \qquad \qquad \qquad \qquad \; 
+ \, p \cdot q_1 \, q_2^{\mu} p^{\nu} 
\, + \, p \cdot q_2 \, p^{\mu} q_1^{\nu} \, \,  )  \nonumber \\
& & \label{eq:mmunu} \\
& & + \; \Frac{C(z,y)}{m_K^2} \; \varepsilon^{\mu \nu \rho \sigma} 
{q_1}_{\rho} {q_2}_{\sigma} \,  
 + \; \Frac{D(z,y)}{m_K^4} \; [ \; \varepsilon^{\mu \nu \rho \sigma}
\, ( p \cdot q_2 \, {q_1}_{\rho} + p \cdot q_1 \, {q_2}_{\rho} ) p_{\sigma}
\nonumber \\
& & \qquad \qquad \qquad \qquad \qquad \qquad \qquad + \, 
( p^{\mu} \varepsilon^{\nu \alpha \beta \gamma} \, + \, p^{\nu} 
\varepsilon^{\mu \alpha \beta \gamma} ) \, p_{\alpha} {q_1}_{\beta} 
{q_2}_{\gamma} \; ] \nonumber
\end{eqnarray}
where
\begin{eqnarray}
y \, = \, \Frac{p \cdot (q_1 - q_2)}{m_K^2} \, & \; \; \; \;  , & \;  
\; \; \; \; z \, = \, \Frac{(q_1 + q_2)^2}{m_K^2}  \; \; \; .
\label{eq:yz}
\end{eqnarray}
The physical region in the adimensional variables $y$ and $z$ is given
by~: 
\begin{eqnarray}
0 \, \leq \, |y| \, \leq \, \Frac{1}{2} \lambda^{1/2} \,
(1,r_{\pi}^2,z) \; & \; 
\; , \; \; & \; 0 \, \leq \, z \leq \, (1-r_{\pi})^2 \; \; \; ,
\label{eq:range}
\end{eqnarray}
with
\begin{eqnarray}
\lambda (a,b,c) & = & a^2 \, + \, b^2 \, + \, c^2 \, - \, 2 \, 
(\, a b \, + \, a c \, + \, b c \, )  \; \; \; ,\nonumber \\
& &  \label{eq:larp} \\
r_{\pi} & = & \Frac{m_{\pi}}{m_K} \; \; \; .\nonumber
\end{eqnarray}
Note that the invariant amplitudes $A(z,y)$, $B(z,y)$ and $C(z,y)$ have
to be symmetric under the interchange of $q_1$ and $q_2$ as required
by Bose symmetry, while $D(z,y)$ is antisymmetric. In the limit where 
CP is conserved the amplitudes $A$ and $B$ contribute only to \kpiggn,
while $C$ and $D$ only contribute to $K_S \rightarrow \pi^{\circ}
\gamma \gamma$. In \kpiggp all of them are involved. 
\par
Using the definitions (\ref{eq:mmunu},\ref{eq:yz}) the double differential
rate for unpolarized photons is given by 
\begin{eqnarray}
\Frac{\partial^2 \Gamma}{\partial y \, \partial z} & \, = \, & 
\, \Frac{m_K}{2^9 \pi^3} \left[ \, z^2 \left( \, | \, A \, + \, B \, |^2
 \, + \, | \, C \, |^2 \right)  \; \right. \label{eq:doudif} \\
& & \; \; \;  \; \; \; \; \; \; \; \left.
+ \, \left( \, y^2 \, - \, \Frac{1}{4} \lambda (1,r_{\pi}^2,z) \, \right)^2
\, \left( \, | \, B \, |^2 \, + \, | \, D \, |^2 \right) \, \right]
\; \; \; . \nonumber
\end{eqnarray}
The processes $K \rightarrow \pi \gamma \gamma$ have no tree level
${\cal O}(p^2)$ contribution. At ${\cal O}(p^4)$ the amplitudes
$B(z,y)$ and $D(z,y)$ are still zero since there are not enough powers
of momenta to generate the gauge structure, and therefore their leading
contribution is ${\cal O}(p^6)$. As can be seen from (\ref{eq:doudif})
only the $B$ and $D$ terms contribute for small $z$ (the invariant
amplitudes are regular in the small $y,z$ region). The antisymmetric
character of the $D(z,y)$ amplitude under the interchange of $q_1$ 
and $q_2$ means effectively that while its leading contribution is 
${\cal O}(p^6)$ this only can come from a finite loop calculation
because the leading counterterms for the $D$ amplitude are 
${\cal O}(p^8)$. However also this loop contribution is helicity suppressed
compared to the $B$ term. As  shown in a similar situation in the 
electric Direct Emission of 
 $K_L \rightarrow \pi^+ \pi^- \gamma$ \cite{DI95},  
this antisymmetric ${\cal O}(p^6)$ loop contribution might be smaller than
the local ${\cal O}(p^8)$ contribution.

\section{\kpiggp at ${\cal O}(p^4)$}
\hspace*{0.5cm}
The leading $\Delta I \, = \, 1/2$ ${\cal O}(p^4)$ $A(z,y)$ and 
$C(z,y)$ amplitudes for \kpiggp have already been computed in 
\cite{EPR88}. We review them here.
\par
The $A$ amplitude reads
\begin{equation}
A^{(4)}(z) \, = \, \Frac{G_8 m_K^2 \alpha_{em}}{2 \pi z} \, \left[ 
\, (z+1-r_{\pi}^2) \, F(\Frac{z}{r_{\pi}^2}) \, + \, 
(z+r_{\pi}^2-1) \, F(z) \, - \, \hat{c} z \, \right] \; \; \; .
\label{eq:atoni}
\end{equation} 
Here $G_8$ is the effective weak coupling constant determined from 
$K \rightarrow \pi \pi$ decays at ${\cal O}(p^2)$
\begin{equation}
| G_8 | \, \simeq \, 9.2 \, \times \, 10^{-6} \, GeV^{-2}
\label{eq:g8fer}
\end{equation}
and the $F(x)$ function is defined as
\begin{eqnarray}
F(x) \, & = & \, \left\{ \begin{array}{ll} 
                     1 \, - \, \Frac{4}{x} \, \arcsin^2 (\Frac{\sqrt{x}}{2})
                     & \; \; , \; \mbox{$x \leq 4$} \nonumber \\
                     &  \nonumber \\
                     1 \, + \, \Frac{1}{x} \, 
                     \left( \ln^2 \left( \Frac{1-\beta(x)}{1+\beta(x)} \right)
                            \, - \, \pi^2 \, + \, 2 i \pi 
                     \ln \left( \Frac{1-\beta(x)}{1+\beta(x)} \right) \right)
                     & \; \; , \; \mbox{$x > 4$} \nonumber 
                     \end{array}
                 \right. \nonumber \\
& & \label{eq:fxbx} \\
\beta(x) \, & \, = \, & \, \sqrt{1-\Frac{4}{x}}  \; \; \; . \nonumber
\end{eqnarray}
In (\ref{eq:atoni}) the pion loop contribution $F(z/r_{\pi}^2)$ dominates
by far over the kaon loop amplitude with $F(z)$. The loop results are 
finite. However as we have already commented $\chi PT$ allows an 
${\cal O}(p^4)$ scale independent local contribution that 
in (\ref{eq:atoni}) is parameterized by 
\begin{equation}
\hat{c} ={ {128 \pi^2}\over{3}}[3(L_9+L_{10}) +N_{14}-N_{15}-2N_{18}] ~,
\label{eq:c-hat}\end{equation}
that is a quantity ${\cal O}(1)$. The $L_9$ and $L_{10}$ are the 
local ${\cal O}(p^4)$
strong couplings and $N_{14},N_{15}$ and $N_{18}$ are ${\cal O}(p^4)$
weak couplings,  still  not fixed by the phenomenology, and that can be  only
computed in a model dependent way \cite{EK93}. The Weak Deformation Model (WDM)
\cite{EP90} predicts $\hat{c}=0$, while naive factorization 
in the Factorization 
Model (FM) \cite{PI91,EK93} gives $\hat{c} = -2.3$. In these models, 
due to the cancellation in the vector meson contribution in (\ref{eq:c-hat}),
the role of axial mesons could be relevant \cite{EK93}.
\par
The ${\cal O}(p^4)$ contribution to the $C(z,y)$ amplitude is 
\begin{equation}
C(z) \, = \, \Frac{G_8 m_K^2 \alpha_{em}}{\pi} \; \left[
             \Frac{z \, - \, r_{\pi}^2}{z \, - \, r_{\pi}^2 \, + \, 
             i r_{\pi} \Frac{\Gamma_{\pi^{\circ}}}{m_K}} \; - \; 
             \Frac{z \, - \, \Frac{2 \, + \, r_{\pi}^2}{3}}{z \, - \, 
              r_{\eta}^2} \right] \; \; \; , 
\label{eq:ctoni}
\end{equation}
where $r_{\eta} = m_{\eta}/m_K$ and $\Gamma_{\pi^{\circ}} \equiv 
\Gamma ( \pi^{\circ} \rightarrow \gamma \gamma )$. This amplitude 
is generated by the Wess--Zumino--Witten functional \cite{WZW71}
$(\pi^{\circ}, \eta) \rightarrow \gamma \gamma$ through the sequence
$K^+ \rightarrow \pi^+ (\pi^{\circ},\eta) \rightarrow \pi^+ \gamma
\gamma$. This contribution amounts roughly to less than $10\%$ in the 
total width.

\section{${\cal O}(p^6)$ local amplitudes for $K^+ \rightarrow
\pi^+ \gamma \gamma$}
\hspace*{0.5cm} 
At ${\cal O}(p^6)$ 
there are only four independent
Lorentz invariant local amplitudes contributing to \kpiggp~:
\begin{eqnarray}
F_{\mu \nu} \, F^{\mu \lambda} \, \partial^{\nu} K^+ \, \partial_{\lambda}
\pi^- \; \; \; \; &  , & \; \; \; \;   
F_{\mu \nu} \, F^{\mu \nu} \, \partial_{\lambda} K^+ \, 
\partial^{\lambda} \pi^- \; \; \; \; , \nonumber \\
& & \label{eq:kind} \\
m_K^2 \, F_{\mu \nu} F^{\mu \nu} \, K^+ \, \pi^- \; \; \; \; & , & \; \; 
\; \;  
\partial^{\alpha} F_{\mu \nu} \, \partial_{\alpha} F^{\mu \nu} \, 
K^+ \pi^- \; \; \; , \nonumber 
\end{eqnarray}
where the last one has no analogous in $K_L \rightarrow \pi^{\circ} 
\gamma \gamma$.
In general the couplings of these operators
 are not directly related with the ones in \kpiggn due to the
fact that the electric charge matrix does not conmute with the 
generators of the electrically charged field.
As we shall see, in specific models, relations between the two channels
can be found.
All of these operators contribute to $A(z,y)$ but only the first one in 
(\ref{eq:kind}) gives a $B(z,y)$ amplitude. 
Loop divergences at ${\cal O}(p^6)$ are absorbed in the counterterm 
coefficients, that thus renormalized are finite.
Chiral dimensional analysis
\cite{WG79} tells us that their contributions are 
suppressed compared to ${\cal O}(p^4)$ by a factor 
$m_K^2/(4\pi F_\pi)^2 \sim 0.2$.
Nevertheless vector meson exchange was found to 
enhance this up to $m_K^2/m_V^2\sim 0.4$ \cite{EG89}.
Thus we  try to estimate the  contributions of
the lightest resonances, i.e. vector mesons, and we assume that
heavier resonances and non--resonant contributions give smaller corrections.
 This picture seems well verified in the  
strong coupling constants at ${\cal O}(p^4)$ \cite{EG89}, and it is 
likely to apply to the weak couplings. 
\par
We are going to consider here the contribution of vector meson 
resonances to the weak ${\cal O}(p^6)$ counterterms for \kpiggp
 (a more complete discussion and an extension to \kpiggn is 
given in \cite{DP96}) and we find that only the terms in (\ref{eq:kind})
with derivatives on the meson fields can be generated by vector meson 
exchange.
\par
Following the study of \kpiggn in \cite{EP90} we assume that the
contribution to the local amplitudes for \kpiggp is dominated by 
vector meson resonances and define their contribution with an 
adimensional parameter $a_V$ generated by the first term in 
(\ref{eq:kind}) as 
\begin{equation}
a_V \, = \, - \, \Frac{\pi}{2 G_8 m_K^2 \alpha_{em}} \, 
\lim_{z \rightarrow 0} \, B_V(z)
\label{eq:avb}
\end{equation}
where $B_V(z)$ is the vector resonance contribution to the $B$ 
amplitude. The $A$ amplitude generated by vector exchange $(A_V)$
gets contributions from the first and second structure in 
(\ref{eq:kind}). However as seen in \cite{EP90} if we assume 
that these local amplitudes are generated through strong resonance
exchange supplemented with a weak transition in the external legs,
those two contributions are related and can be written in terms of the
$a_V$ parameter defined in (\ref{eq:avb}) as
\begin{equation}
A_V \; = \; \Frac{G_8 m_K^2 \alpha_{em}}{\pi} \, a_V \, 
(\, 3 \, + \, r_{\pi}^2 \, - \, z \, )
\label{ampvb}
\end{equation}
In \cite{EP90} two different vector contributions to $a_V$ in 
(\ref{eq:avb}) have been proposed : the first one amounts for the
weak counterterms generated by a strong vector resonance exchange
with a weak transition in an external leg ($a_V^{ext}$), the second
is generated by vector resonance exchange between a direct weak 
vector--pseudoscalar--photon ($VP\gamma$) vertex and a strong one 
($a_V^{dir}$). While the 
first can be computed in a model independent way, the generation
of direct $VP\gamma$  weak vertices is still poorly known and therefore only 
models can be used.
\par
The {\em external} weak transition for \kpiggp gives
\begin{equation}
a_V^{ext} \, = \, - \, \Frac{128 \pi^2 \, h_V^2 \, m_K^2}{9 \, m_V^2} 
\; = \; -0.08
\label{eq:avext}
\end{equation}
where $m_V = m_{\rho}$ and the strong 
$VP\gamma$ coupling $|h_V| = (3.7 \pm 0.3) \times 10^{-2}$ is defined
by
\begin{equation}
{\cal L}(VP\gamma) \, = \, h_V \, \varepsilon_{\mu \nu \rho \sigma}
 \, \langle \, V^{\mu} \, \{ \, u^{\nu} \, , \, f_+^{\rho \sigma} \, \}
\rangle \; \; \; , 
\label{eq:vpga}
\end{equation}
with
\begin{eqnarray}
u_{\mu} \, = \, i u^{\dagger} D_{\mu} U  u^{\dagger} & \; \; \;  , \; \; \; &
f_+^{\mu \nu} \, = \, u F_L^{\mu \nu} u^{\dagger} \, + \, 
u^{\dagger} F_R^{\mu \nu} u \; \; \; , \nonumber \\
& & \nonumber \\
U \, = \, u u  & \; \; \; \; \; \; \; \; , \; \; \; 
& D_{\mu} U \, = \, \partial_{\mu} U \, 
- \, i r_{\mu} U \, + \, i U \ell_{\mu} \; \; \; , \label{eq:nota} \\
& & \nonumber \\
U & = & \exp \left( \, \Frac{i}{F_{\pi}}\, \sum_{j=1}^{8} \, \lambda_j
\, \phi_j \, \right) \; \; \; , 
\nonumber 
\end{eqnarray}
and $u(\phi)$ is an element of the coset space $SU(3)_L \otimes 
SU(3)_R \, / \, SU(3)_V$ parameterized in terms of the Goldstone
fields $\phi_i$ $i=1,...8$, $F_{R,L}^{\mu \nu}$ are the strength field
tensors associated to the external $r_{\mu}$ and $\ell_{\mu}$ fields,
$V_{\mu}$ is the nonet of vector meson resonances, $F_{\pi} \, \simeq \, 
93 \, MeV$ is the decay constant of pion 
and, finally, $\langle A \rangle \, \equiv \, Tr(A)$. In our case
with two photon external fields $\ell_{\mu} = r_{\mu} = \, e Q A_{\mu}$~
\footnote{For a full discussion of the definitions and notations see
\cite{EG89,EP90}.} .
\par
The model dependent contribution to $a_V$ from {\em direct}
weak vertices in the WDM and FM quoted above is~:
\begin{eqnarray} 
a_{V \, , \, WDM}^{dir}  \, &  =  & \, - \, a_V^{ext} \; \; \; , \nonumber \\
& & \label{eq:avdir} \\
a_{V \, , \, FM}^{dir}  \, & = & - \, 2 k_F \, a_V^{ext} \; \; \; , \nonumber
\end{eqnarray}
where in the FM $k_F$ is the unknown fudge factor that is not fixed
by the model and satisfies $0 \, < \, k_F \, \leq 1$. Naive 
factorization predicts $k_F \, = \, 1$~.
\par
By adding (\ref{eq:avext}) and (\ref{eq:avdir}) we see that 
$a_V \, = \, a_V^{ext} \, + \, a_V^{dir}$ is 
\begin{eqnarray}
a_{V \, , \,  WDM} \, & = & \, 0 \; \; \; \; , \nonumber \\
& & \label{eq:avtot} \\
a_{V \, , \,  FM} \, & = & \, (1 - 2 k_F) \, a_V^{ext} \; \; \; \; .
\nonumber
\end{eqnarray}
We note that for the allowed range of values of $k_F$ in the FM both 
models agree in predicting a very small vector meson contribution to
${\cal O}(p^6)$ in \kpiggp and therefore
the important conclusion of our exercise is that the vector contribution
to the local $B(z,y)$ amplitude can be neglected. If, as it is 
reasonable to 
assume, other resonance interchange corrections are even smaller, 
we can conclude that
the small $z$ region of \kpiggp is completely predictable at 
${\cal O}(p^6)$ through chiral loops and it is likely to be dominated 
by the unitarity corrections of 
$K^+ \rightarrow \pi^+ \pi^+ \pi^-$ to \kpiggp. Indeed in that region
the ${\cal O}(p^6)$ is dominant and our ignorance on the
$\hat{c}$ amplitude is irrelevant.

\section{Unitarity corrections of $K^+ \rightarrow \pi^+ \pi^+ 
\pi^-$ to \kpiggp}
\hspace*{0.5cm}
The amplitude for the process $K(p) \rightarrow \pi(p_1) \pi(p_2)
\pi(p_3)$ can be expanded in powers of the Dalitz plot variables
\begin{eqnarray}
X \, = \, \Frac{s_2 - s_1}{m_{\pi}^2} \; \; & , & \; \; \; 
Y \, = \, \Frac{s_3 - s_{\circ}}{m_{\pi}^2} \; \; \; , 
\label{eq:xy}
\end{eqnarray}
where $s_i = (p - p_i)^2$ for $i=1,2,3$ , $s_{\circ} =
(s_1 + s_2 + s_3)/3$ and the subscript $3$ indicates the odd pion. 
For the decay
$K^+(p) \rightarrow \pi^+(p_1) \pi^+(p_2) \pi^-(p_3)$ the isospin
decomposition, neglecting the small phase shifts and up to quadratic
terms, is written as \cite{ZE64,KM91}
\begin{eqnarray}
A(K^+ \rightarrow \pi^+ \pi^+ \pi^-) \, & = & \,
2 \, \alpha_1 \, - \, \alpha_3 \, + \, ( \, \beta_1 \, - \, 
\Frac{1}{2} \beta_3 \, + \, \sqrt{3} \gamma_3 \, ) \, Y  \label{eq:ay2x2} \\ 
& & \, - \, 2 \, ( \, \zeta_1 \, + \, \zeta_3 \, ) \, ( \, Y^2 \, 
+ \, \Frac{X^2}{3} \, ) \, - \, ( \, \xi_1 \, + \, \xi_3 \, - 
\, \xi_3' \, ) \, ( \, Y^2 \, - \, \Frac{X^2}{3} \, ) \; , \nonumber
\end{eqnarray}
where the subscripts $1$ and $3$ refer to $\Delta I \, = \, 1/2 \, , \, 3/2$
transitions respectively, and the coefficients in (\ref{eq:ay2x2}) have been 
fitted to the data  \cite{KM91}. The ${\cal O}(p^2)$ amplitude 
\begin{equation}
A^{(2)} (K^+ \rightarrow \pi^+ \pi^+ \pi^-) \, = \, G_8 m_K^2 \, 
\left( \Frac{2}{3} \, - \, r_{\pi}^2 \, Y \, \right) 
\label{eq:ap2k}
\end{equation}
with the value of $G_8$  in (\ref{eq:g8fer}) 
underestimates by 20-30\%
 the experimental linear slopes for $K^+ \rightarrow \pi^+ \pi^+ \pi^-$. 
At the next chiral order, the  experimental linear and quadratic slopes in
(\ref{eq:ay2x2}) are recovered  (with predictive power too) \cite{KM91}.  
Since the ${\cal O}(p^4)$ loop contribution  (\ref{eq:atoni}) to
\kpiggp, is generated by (\ref{eq:ap2k}),
it seems natural to try to include all the contributions to \kpiggp 
generated by the experimental slopes in (\ref{eq:ay2x2}),
i.e. we evaluate the  ${\cal O}(p^6)$ contributions to \kpiggp 
induced by the ${\cal O}(p^4)$ corrections to
$K^+ \rightarrow \pi^+ \pi^+ \pi^-$.
This can be done similarly to the case of \kpiggn in \cite{CD93,CE93}:
(\ref{eq:ay2x2}) is considered  as an effective  $K^+ \rightarrow \pi^+ \pi^+
\pi^-$  chiral vertex,  the kinematical  variables  are replaced by the
appropriate covariant derivatives, the QED scalar vertices are added 
through minimal coupling and then the 
usual Feynman diagrams approach can be used. There are 13 Feynman diagrams 
that have two different topologies. There
is a subset of 9 diagrams where one or both photons are radiated
by the external legs and therefore do not give and absorptive
contribution. The sum of these
bremsstrahlung--like diagrams is not gauge invariant and cancels with an
analogous term from the  4 remaining diagrams (two of which have also an 
absorptive part) and generate the
$A(z,y)$ and $B(z,y)$ amplitudes. 
This loop result for the $A(z,y)$ and $B(z,y)$ amplitudes is divergent
and needs to be regularized and renormalized. We have used dimensional
regularization and  divergences are absorbed by the four 
counterterms discussed earlier.
\par
Alternatively one could use a subtracted dispersion relation.
While the absorptive contribution 
is uniquely determined, the dispersive one, due to the presence of
subtraction constants, can be computed only up to a polynomial.
This is precisely the same ambiguity that happens in the effective
vertex method that we have used where the finite part of the 
counterterms is a priori unknown.
\par
We assume that the dominant contribution should come from the 
non--polynomial amplitude generated by the cut of the two--pion
intermediate state. Otherwise contributions generated by a vanishing
on--shell $K \rightarrow 3 \pi$ amplitude can be reabsorbed in the
unknown polynomial amplitude.
\par
The final result for the $A$ and $B$ amplitudes for \kpiggp in the
$\overline{MS}$ subtraction scheme are

\begin{eqnarray}
A(z,y)  & = &   \, \Frac{\alpha_{em}}{2 \pi} \, \cdot 
\nonumber \\
& &  \left\{ \, \left[ \, 2 ( 2 \alpha_1 - \alpha_3) \, + \, 
\left( 1 + \Frac{1}{3 r_{\pi}^2} - \Frac{z}{r_{\pi}^2} \right) \, 
\left( \beta_1 - \Frac{1}{2} \beta_3 + \sqrt{3} \gamma_3 \right) \, 
\right] \, \Frac{1}{z} \, F \left( \Frac{z}{r_{\pi}^2} \right) \right.
\nonumber \\
& & \; \; \left. - \, \Frac{8}{3 r_{\pi}^4} \, ( 2 \zeta_1 - \xi_1 ) 
\left[ 
\, r_{\pi}^2 \, \left( \ln \left( \Frac{m_{\pi}^2}{\mu^2}
\right) 
\, - \, 1 \, \right) \, \right. \right. \nonumber \\
& & \qquad \qquad \qquad \; \; \; \; \; \; \; \left. \left.
+ \, \Frac{1}{18} \, ( 1 + 6 (r_{\pi}^2 - z) + 9 (r_{\pi}^2-z)^2 )\, 
\Frac{1}{z}
F \left( \Frac{z}{r_{\pi}^2} \right)  \right] \right. \nonumber \\
& & \; \; \left. 
- \, \Frac{8}{3 r_{\pi}^4} \, ( 4 \zeta_1 + \xi_1) 
\left[ \, 
 - \Frac{1}{12} ( 1 + 6 r_{\pi}^2) 
 \ln \left( \Frac{m_{\pi}^2}{\mu^2} \right) \, + \, \Frac{r_{\pi}^2}{2} 
\right. \right. \, \nonumber \\
& & \qquad  \qquad \; \; \; \; \; \; \; \; \; \; \; \; \; \; \;   \left. \left. 
- \, \Frac{1}{36} \left( 9 r_{\pi}^2 - 5 - 3(1+3 r_{\pi}^2) (r_{\pi}^2-z)
\right) \Frac{1}{z} F \left( \Frac{z}{r_{\pi}^2} \right) \, \right. \right.
\nonumber \\
& & \qquad \qquad \; \; \; \; \; \; \; \; \; \; \; \; \; \;   \left. \left. 
\, + \Frac{y^2}{z} \left( \Frac{1}{12}
\, + \, 3 R \left( \Frac{z}{r_{\pi}^2} \right) \, + \, \Frac{1}{2} \left( 1 + 
\Frac{2 r_{\pi}^2}{z} \right) \, F \left( \Frac{z}{r_{\pi}^2} \right) 
\right) \right. 
\right.
\nonumber \\
& &  \qquad \qquad \qquad \; \; \; \; \; \; \; \; \left. \left. 
- \Frac{(1-r_{\pi}^2+z)^2}{4 z} \left(
\Frac{1}{12} \, + \, R \left( \Frac{z}{r_{\pi}^2} \right) \, + \,
\, \Frac{1}{2} \left( 1 + 
\Frac{2 r_{\pi}^2}{z} \right) \, F \left( \Frac{z}{r_{\pi}^2} \right) 
\right) \right. 
\right. \nonumber \\
& & \qquad \qquad \qquad \; \; \; \; \; \; \; \; \left. \left. + 
(1-r_{\pi}^2+z) \, \left( 
\Frac{1}{24} \, + \, \Frac{z}{72 r_{\pi}^2} \, \right. \right. \right. 
\nonumber \\ 
& & \qquad \qquad \qquad \qquad \qquad \qquad \; \; \; \; \; \; \; \; \;  
\left. \left. \left.
\,  + \, 
\Frac{1}{12}  \left( 1 + \Frac{2 r_{\pi}^2}{z} \right) 
\left( \Frac{z}{r_{\pi}^2} R \left( \Frac{z}{r_{\pi}^2} \right) + 3 
F \left( \Frac{z}{r_{\pi}^2} \right) \right) \right)  \right. \right. 
\nonumber \\
& & \qquad \qquad \qquad  \; \; \; \; \; \; \; \; \; \left. \left. - z  
\left( \Frac{1}{36} - 
\Frac{r_{\pi}^2}{24 z} + \Frac{z}{72 r_{\pi}^2} + 
\Frac{1}{12} \left( \Frac{z}{r_{\pi}^2} + 1 - \Frac{2 r_{\pi}^2}{z} 
\right) R \left( \Frac{z}{r_{\pi}^2} \right) \right. \right. \right. 
\nonumber \\
& & \qquad \qquad \qquad \; \; \; \; \; \; \; \; \; \; \; \; \; \; \,  
\left. \left. \left. 
+ \Frac{r_{\pi}^2}{2 z}
\left( 1 - \Frac{r_{\pi}^2}{z} \right) F \left( \Frac{z}{r_{\pi}^2} \right)
\right) \right] \right. \nonumber \\
& & \; \; \;  \left.  
+ \, G_8 m_K^2 \left[ \, (z + r_{\pi}^2 - 1) \Frac{1}{z} F(z) \, - 
\, \hat{c} \, + \, 2 r_{\pi}^2 \eta_1 \, + \, 2 \eta_2 \, \right] 
\right\} \; \; \; \; , 
\label{eq:ayz6}
\end{eqnarray}
\newpage
\begin{eqnarray}
B(z,y) &  = &  \Frac{\alpha_{em}}{\pi}\, \left\{ \, 
               \Frac{1}{3 r_{\pi}^4} ( 4 \zeta_1 + \xi_1)  \, 
               \left[  - \, \Frac{1}{6} \, \left( \, 1 \, + \, 2 \, 
\ln \left( \Frac{m_{\pi}^2}{\mu^2} \right) \, \right) \, + \, 
\Frac{z}{18 r_{\pi}^2} \, 
- \, \Frac{2 r_{\pi}^2}{z} \, F \left( \Frac{z}{r_{\pi}^2} \right) \, 
\right. \right. \nonumber \\
& & \qquad \qquad \qquad \; \; \; \; \; \; \; \;  \; \; \;  
\left. \left. + \, \Frac{1}{3} \, \left( \, 
\Frac{z}{r_{\pi}^2} \, - \, 10 \right) \, R \left( \Frac{z}{r_{\pi}^2} 
\right) \, \right] \right. \,\nonumber \\
& & \left. \qquad \; \;  + \, G_8 m_K^2 \eta_3 \, \right\} \; \; \; \; , 
\label{eq:byz6}
\end{eqnarray}

where the $F(x)$ function has been defined in (\ref{eq:fxbx}) and

\begin{equation}
R(x) \, = \, \left\{ \begin{array}{ll}
                    - \, \Frac{1}{6} \, + \, \Frac{2}{x} \, - \,
                    \Frac{2}{x} \, \sqrt{\Frac{4}{x} \, - \, 1} \, 
                    \arcsin ( \Frac{\sqrt{x}}{2} ) &  
                    , \; \mbox{$x \leq 4$} \\
                    & \\     
                    - \, \Frac{1}{6} \, + \, \Frac{2}{x} \, + 
                    \, \Frac{\beta (x)}{x} \, \left( \, \ln \left(
\Frac{1\, - \, \beta (x)}{1 \, + \, \beta (x)} \right) \, 
+ \, i \pi \, \right)
& , \; \mbox{$x > 4$} \end{array} \right.
 \label{eq:rx} 
\end{equation}
with $\beta (x)$ defined in (\ref{eq:fxbx}) ($F(z) \simeq - \, 
\Frac{z}{12}$ and $ R(z) \simeq \Frac{z}{60}$ for $z \rightarrow 0$ ) . 
\par
In (\ref{eq:ayz6},\ref{eq:byz6}) the $\eta_i, \, i=1,2,3$ stand for
the unknown polynomial contribution we have spoken before. 
Though, in principle, four different subtraction constants could appear,
these are the only ones that are necessary in order to absorb the 
loop divergences generated by this unitarity correction.
We emphasize
that these local amplitudes are not generated by the vector mesons
(already taken into account by $a_V$) and therefore are expected to be
suppressed by $m_K^2/\Lambda_{\chi}^2$ with $\Lambda_{\chi} \simeq
4 \pi F_{\pi}$ over the previous chiral order. If as a naive chiral
dimensional analysis we choose as coefficients of the suppression factor
the factor accompanying $\ln (m_{\pi}^2/\mu^2)$ (for definiteness) we get
\begin{equation}
\eta_1 \, = \, 12 \Frac{m_K^2}{\Lambda_{\chi}^2} \simeq 2.06 
\qquad , \; \; \; \; 
\eta_2 \, = \, \Frac{7}{5} \Frac{m_K^2}{\Lambda_{\chi}^2} \simeq 0.24
\qquad , \; \; \; \; 
\eta_3 \, = \, - \, \Frac{3}{2} \Frac{m_K^2}{\Lambda_{\chi}^2} \simeq -0.26
\; \; . \label{eq:etasi}
\end{equation}
We note the big numerical value for $\eta_1$, but looking into the 
$A$ amplitude (\ref{eq:ayz6}) we see that $\eta_1$ is 
suppressed by $m_{\pi}^2/m_K^2$. The order of magnitude we get indeed
is the one expected by chiral counting.
In (\ref{eq:ayz6},\ref{eq:byz6}) we only have included the $\Delta I = 
3/2$ coefficients of the ${\cal O}(p^2)$ contribution of 
$K^+ \rightarrow \pi^+ \pi^+ \pi^-$  because the bigger errors in the
determination of these coefficients in the ${\cal O}(p^4)$ amplitudes.
\par
We notice that at this order while the 
$B$ amplitude is $y$--independent the $A$ amplitude gets its first
$y$ dependence. 
\par
As we have seen the vector meson contribution to the $B$ amplitude
is likely to be very small. If we use the same definition (\ref{eq:avb})
for our $B$ amplitude (\ref{eq:byz6}) we get
\begin{equation}
a_{\ell} \, = \, \Frac{4 \zeta_1 + \xi_1}{18 G_8 m_K^2 r_{\pi}^4} \, 
\ln \left( \Frac{m_{\pi}^2}{\mu^2} \right) \, - \, \Frac{1}{2} \eta_3
\; \; \; \; .
\label{eq:all}
\end{equation}
The first term amounts to $\simeq 0.52$ if $\mu = m_{\rho}$ is taken.
Then from (\ref{eq:etasi}) we can conclude that due to the expected
vanishing of the vector meson contributions, 
the polynomial non--resonant amplitude, though small compared
with the unitarity corrections, could be in principle tested 
in this channel at low z. We
note that this situation is different to the \kpiggn case where
the vector meson contribution seems to be more relevant.
\par
We have not included the $\Delta I = 3/2$ couplings in the 
${\cal O}(p^6)$ contribution because of the big errors in the experimental 
fit. But we should stress that inputting those amplitudes could modify
the low--z behaviour of the $B$ amplitude even a $30\%$.
\par
The argument about the relevance of the correction to 
\kpiggn coming from the inclusion
of the experimental $\gamma \gamma \rightarrow \pi^{\circ} \pi^{\circ}$
amplitude \cite{KH94} is weakened here due to the relative suppression of 
$K^+ \rightarrow \pi^+ \pi^{\circ} \pi^{\circ}$ compared to
$K^+ \rightarrow \pi^+ \pi^+ \pi^-$.

\section{\kpiggp: branching ratio and spectra}
\hspace{0.5cm}
We can now proceed to study the numerical results for \kpiggp taking
into account the ${\cal O}(p^4)$ and our ${\cal O}(p^6)$ loop 
evaluation of the unitarity corrections. As we have shown the  
${\cal O}(p^6)$ vector resonance dominated local contributions are 
negligible. In our numerical discussion we will input $\eta_i = 0, \, 
i=1,2,3$ and $\mu = m_{\rho}$.
\par
At present there is only an upper bound for Br($K^+ \rightarrow
\pi^+ \gamma \gamma$) which depends
on the shape of the spectrum \cite{AT90}
\begin{eqnarray}
Br(\kpiggpe) \, \left. \vert_{exp} \right. \,  
& \leq & \, 1.5 \times 10^{-4} \; \; \; 
\; \; \; (\chi PT \; amplitude) \; \; \; ,  \nonumber \\
& & \label{eq:brexp} \\
Br(\kpiggpe) \, \left. \vert_{exp} \right. \,  
& \leq & \, 1.0 \times 10^{-6}  \; \; \; 
\; \; \; (constant \; amplitude) \; \; \; \; . \nonumber 
\end{eqnarray}
The uncertainty in the theoretical prediction is dominated by the unknown
${\cal O} (p^4)$ counterterm generated amplitude $\hat{c}$ in 
(\ref{eq:atoni}). In Fig. 1 we show Br($K^+ \rightarrow \pi^+ \gamma
\gamma$) as a function of 
$\hat{c}$ with and without the ${\cal O}(p^6)$ corrections that we have
computed. We remind that WDM predicts $\hat{c}=0$ while naive FM gives
$\hat{c}=-2.3$ with the following results~:
\begin{eqnarray}
Br(\kpiggpe) \, \left. \vert_{WDM} \; 
\right. & = & \; 7.24 \times 10^{-7} \; \; \; , 
 \nonumber \\
& & \label{eq:modpre} \\
Br(\kpiggpe) \, \left. \vert_{nFM} \;  \right.  & = & \; 6.20 \times 1
0^{-7} \; \; \; \; .
\nonumber
\end{eqnarray}
When comparing with the ${\cal O}(p^4)$ predictions \cite{EPR88}
we find that the unitarity corrections increase around $30-40\%$ 
the branching ratio.
In Fig. 2 we show the z--distribution  at 
${\cal O}(p^4)$ and with our ${\cal O}(p^6)$ unitarity correction for 
$\hat{c}=0$. As can be seen the correction is noticeable.
\par
The uncertainty on the $\hat{c}$ amplitude also translates into the
spectra. In Fig. 3 we show the spectrum of the invariant mass of the 
two photons for the two values of $\hat{c}$ predicted by the WDM and
naive FM. We notice that the main dependence on $\hat{c}$ arises in the
$z$-region where there is the bulk of the absorptive contribution.
\par
Finally in Fig. 4 we show the spectrum in the ``asymmetric" $y$ variable
for $\hat{c}=0$. At this order this spectrum has much less structure
than the one in the di--photon invariant mass.

\section{Conclusion}
\hspace{0.5cm} 
The situation of the chiral prediction for \kpiggn and the expected 
experimental measurement of $\Gamma(K^+ \rightarrow \pi^+ \gamma
\gamma)$ at DA$\Phi$NE and BNL
 make interesting to enlarge our theoretical knowledge about
this last process.
\par
We have reviewed and studied the vector resonance dominated 
${\cal O}(p^6)$ local amplitudes which main role is to determine
the low--z region of the invariant di--photon mass spectrum. The
conclusion is that this contribution is likely to be negligible.
The ${\cal O}(p^6)$ unitarity corrections from $K^+ \rightarrow
\pi^+ \pi^+ \pi^-$ to \kpiggp have been computed and found 
to be relevant increasing the branching ratio (for a fixed value
of the $\hat{c}$ amplitude) around $30-40\%$. The shape of the
z--distribution is shown to be sensitive to the evaluated 
corrections (Fig. 2 and Fig. 3).
\par
In Fig. 1 we have shown the 
branching ratio for \kpiggp as a function of $\hat{c}$. 
The included corrections will allow to get a more accurate determination of 
$\hat{c}$ once the branching ratio of \kpiggp is measured and therefore
will provide a new independent relation between ${\cal O}(p^4)$ weak
coupling constants that will improve our predictive power in this 
sector.

\vspace*{1cm} 
{\large \bf Acknowledgements}
\vspace*{0.4cm} \\
\hspace*{0.5cm}
The authors thank G. Ecker for interesting correspondence about the 
subject and
the Particle Theory Group at MIT where part of this
work was done. J.P. also thanks the hospitality of the Particle Theory Group at 
Rutherford--Appleton Laboratory where this work was started.
J.P. is supported by an INFN Postdoctoral fellowship and 
 also partially supported by DGICYT under grant PB94--0080.

\newpage

\noindent {\large \bf Figure captions}
\\
\\
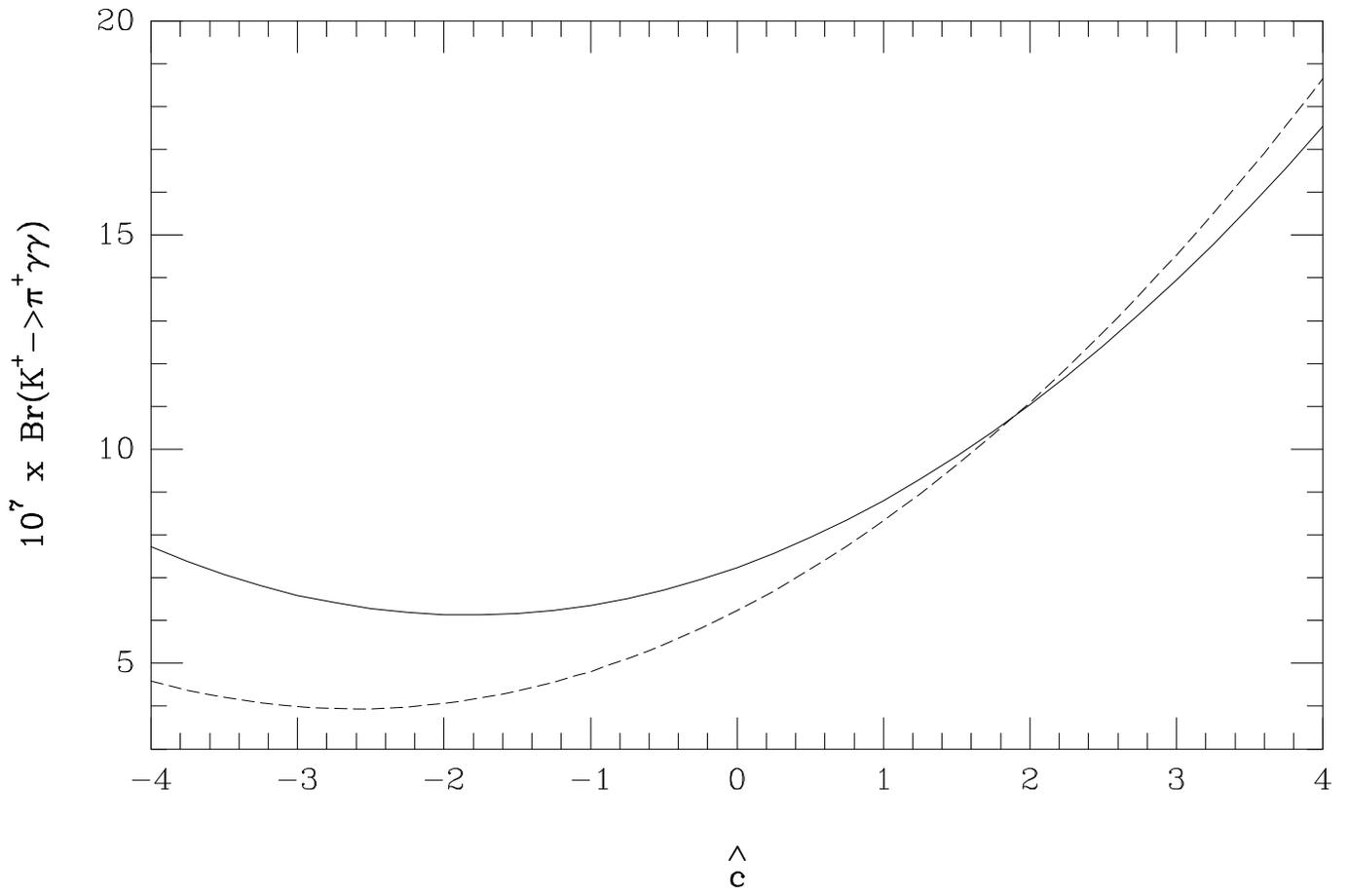
{\bf Fig. 1}~: $Br(\kpiggpe)$ as a function of the $\hat{c}$ amplitude.
The dashed line corresponds to ${\cal O}(p^4)$ $\chi$PT amplitude. The
full line corresponds to the amplitude including the  
${\cal O}(p^6)$ unitarity corrections.
\\
\\
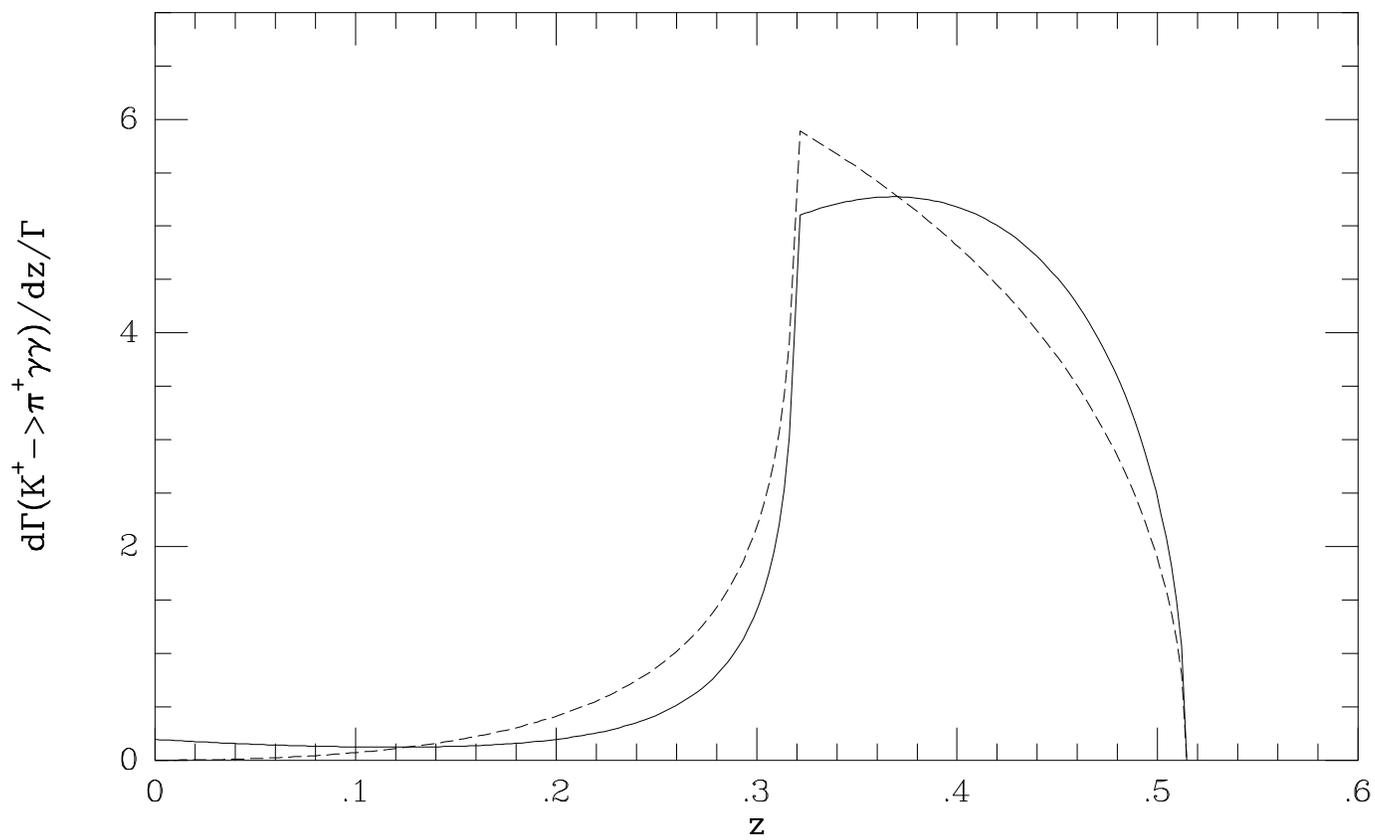
{\bf Fig. 2}~: Comparison of the normalized z--distribution for
\kpiggp at ${\cal O}(p^4)$ (dashed line) with our ${\cal O}(p^6)$ 
correction (full line) for $\hat{c}=0$.
\\
\\
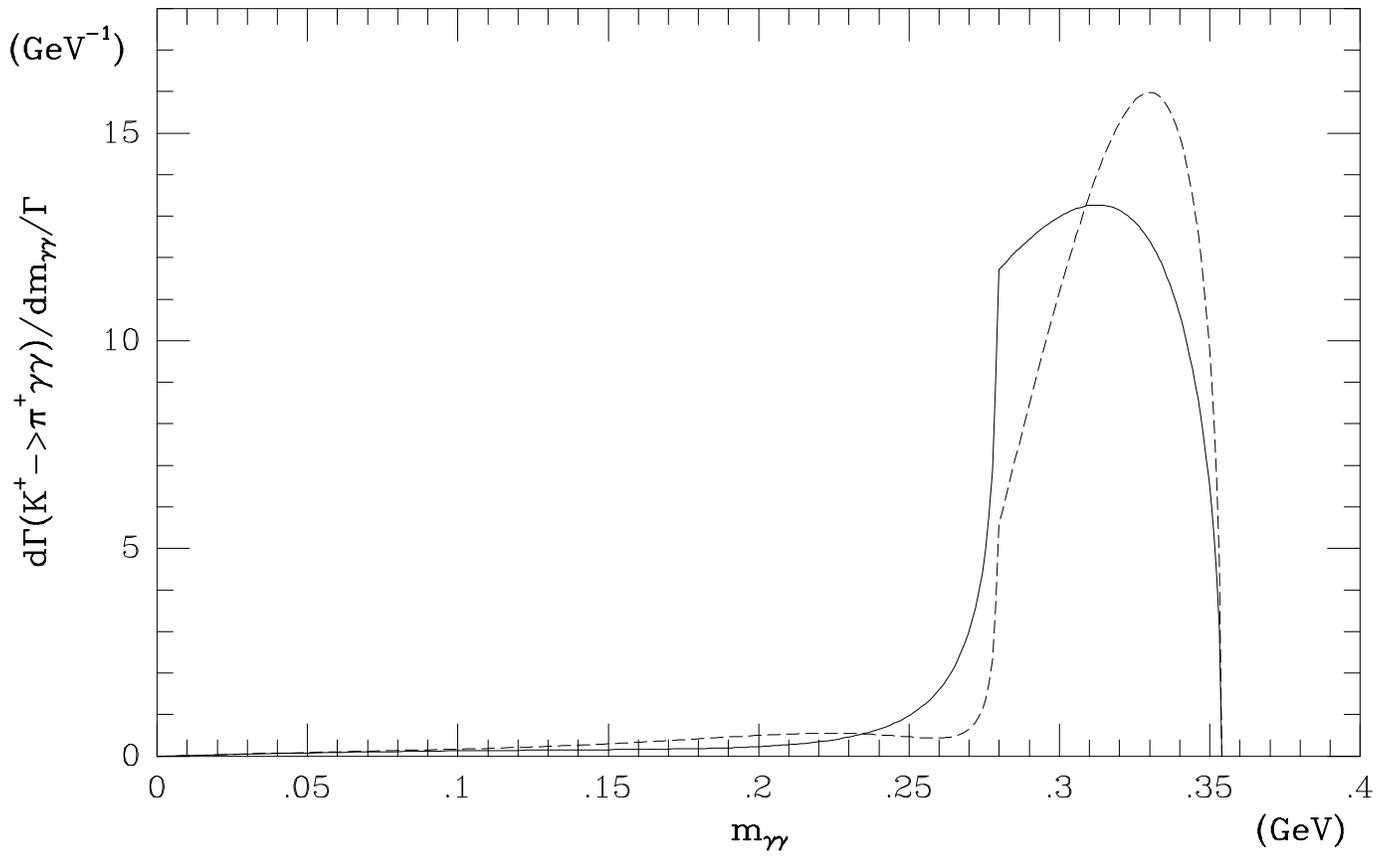
{\bf Fig. 3}~: Normalized di--photon mass spectrum of \kpiggp for
$\hat{c}=-2.3$ (naive FM, dashed line) and $\hat{c}=0$ (WDM , full line).
\\
\\
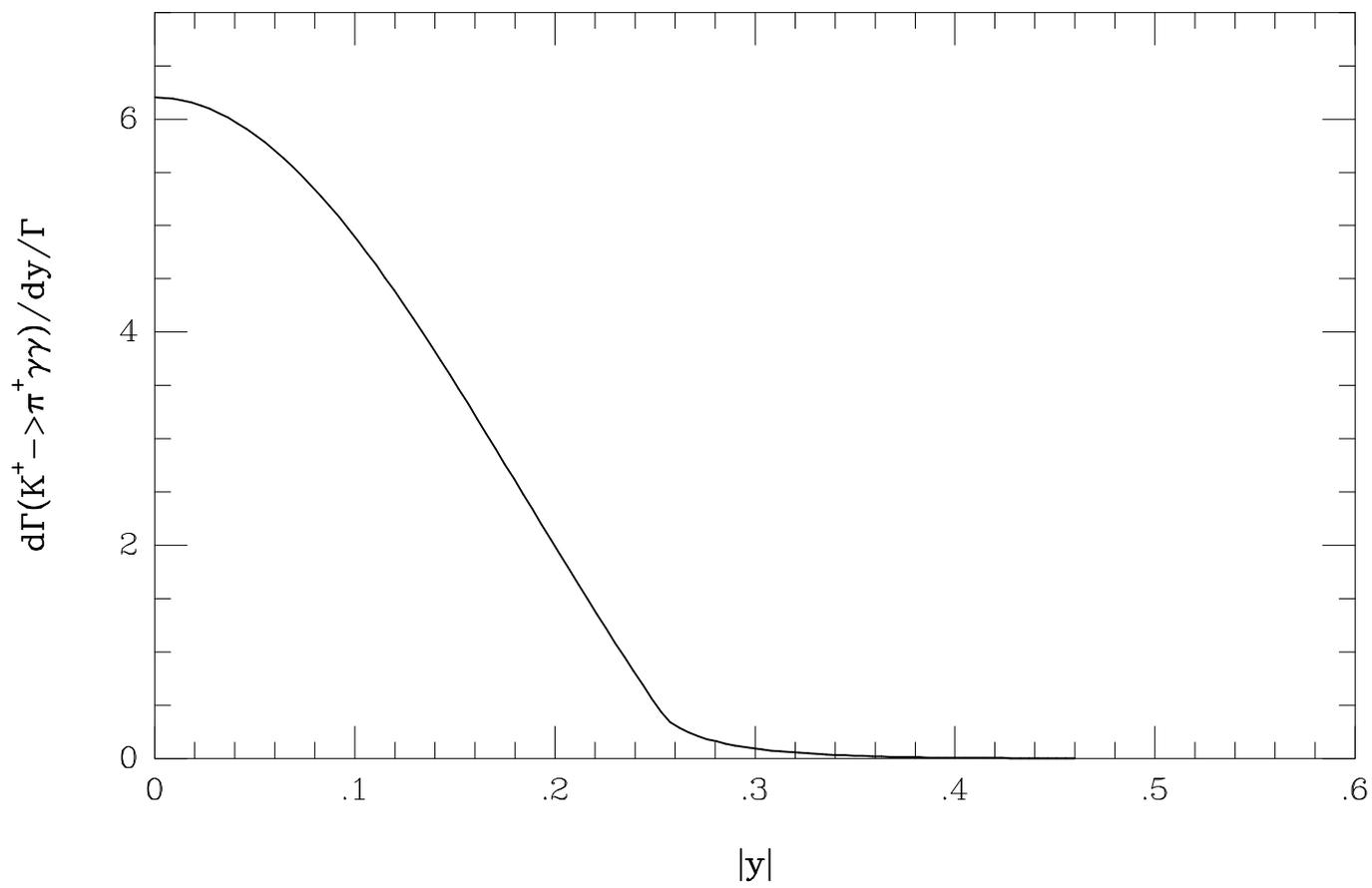
{\bf Fig. 4}~: $\Frac{\partial \, Br(\kpiggpe)}{\partial y}$ 
spectrum for $\hat{c}=0$ as a function of $|y|$.

\end{document}